\documentstyle[pre,aps,epsf,multicol]{revtex}
\draft

\begin{document}
\title{Deviations from linear theory for fluctuations below the supercritical primary bifurcation to electroconvection}
\author{Michael A. Scherer and Guenter Ahlers}
\address{Department of Physics and Quantum Institute,  
University of California, Santa Barbara, California 93106}
\author{Frank H\"orner and Ingo Rehberg}
\address{Experimentalphysik V, 
Universit\"at Bayreuth, D-95440 Bayreuth, Germany}
\date{\today}
\maketitle
\begin{abstract}

We report measurements of thermally-induced mean-square director-angle fluctuations $\langle\theta^2\rangle$ below primary supercritical  bifurcations to electroconvection of the nematic liquid crystals I52 and ``Merck Phase V". For $\epsilon_{\rm mf} \equiv V^2/V_{\rm c,mf}^2 - 1 \alt -0.1$ ($V$ is the amplitude of the applied alternating voltage) we find $\langle\theta^2\rangle \propto |\epsilon_{\rm mf}|^{-\gamma}$ with $\gamma$ given by linear theory (LT). Closer to the bifurcation there are deviations from LT with a smaller $\gamma$ and  
with $V_c^2 > V_{\rm c,mf}^2$. For I52 measurements as a function of the conductivity $\sigma$ of $\langle\theta^2\rangle$ above $V_c^2$ suggest a tricritical bifurcation at $\sigma_t \simeq 4.0\times 10^{-9} \Omega^{-1}{\rm m}^{-1}$, and $(V_c^2 - V_{\rm c,mf}^2)/V_c^2$ increases strongly as $\sigma_t$ is approached from above. 

\end{abstract}

\begin{multicols}{2}

Convection in fluids is a classical system for the study of pattern formation under nonequilibrium conditions because of its similarity to numerous pattern-forming phenomena in nature\ \cite{CH93}. Rayleigh-B\'{e}nard convection (RBC) is the most prominent example. Here a horizontal fluid layer, confined at the bottom and top, is heated from below. If the temperature difference exceeds a threshold, the system undergoes a sharp but continuous (supercritical) bifurcation from a uniform state to a state where patterns, e.g., stripes, squares, or spirals, occur\ \cite{BPA99}. However, already below this bifurcation there exist fluctuations $\delta T$ of the temperature field which become ``large" near the bifurcation point and which are induced by thermal noise. The fluctuation amplitudes have zero mean but a finite mean-square $\langle\delta T^2\rangle$.
For RBC quantitative predictions of $\langle\delta T^2\rangle$ were made three decades ago \ \cite{ZS70,Gr74,Gr75,SH77} on the basis of linear theory (LT) which neglects interactions between the fluctuations. Since $\langle\delta T^2\rangle$ is extremely small, quantitative experimental verifications could be obtained only much more recently\ \cite{QR95,WAC95}.

Closer to the bifurcation, LT should break down because of nonlinear interactions between the fluctuations. In analogy to critical phenomena in equilibrium systems, one then expects a modified ``critical" (rather than ``mean field") behavior of the system. For RBC it was predicted by Swift and Hohenberg\cite{SH77} that the fluctuation interactions should lead to a first-order transition, i.e., to a subcritical bifurcation. Experimentally this interesting phenomenon has been out of reach so far because under most circumstances it is expected to become noticeable only within a few parts per million of the bifurcation point\ \cite{SH77,FN}.

A more favorable system to study the influence of thermal noise in pattern-forming systems is electroconvection (EC) in a nematic liquid crystal (NLC)\ \cite{KP94}. Here an alternating voltage of amplitude $V$ is applied to a cell filled with NLC. The role of the temperature difference is now taken by $V$, and EC occurs for $V > V_c$. For NLCs the effects of thermal noise are larger than for RBC because of the small elastic constants of the NLC\ \cite{FN0}. This already large susceptibility is enhanced even further in EC by the very small thickness of the cell. Indeed, already a decade ago it was possible to visualize the fluctuating convective patches and to measure their amplitudes below onset\ \cite{RRDSHAB91}. Good agreement with LT was found. We do not know of a detailed nonlinear theory for this system. Its critical behavior need not be the same as that of RBC because the anisotropy of the NLC suggests that it may belong to a different universality class. Although one expects the critical region to be wider than for RBC, its study had been beyond experimental resolution heretofore.

We report results for thermally-induced fluctuations of the director angle $\theta$ of the NLCs I52 and ``Merck Phase V" (MPV) both far below and very close to the primary supercritical bifurcation. For I52 we measured the mean square $\langle\theta^2\rangle$, averaged over time as well as over all fluctuating modes of the system as described in\ \cite{BA98,BA99}. In the case of MPV, however, we measured the time averaged mean square amplitudes $\langle A^2 \rangle$ of the fluctuations of only the $x$-component of the critical mode of the system, as described in\ \cite{RHCRW91}. In agreement with earlier work, \cite{RRDSHAB91,BA98,BA99} our data well below $V_c$ are consistent with $\langle\theta^2\rangle \propto \epsilon_{\rm mf}^{-\gamma}$ ($\langle A^2 \rangle \propto \epsilon_{\rm mf} ^{-\gamma_0}$)  with $\gamma = 1/2$ ($\gamma_0 = 1$) as predicted by LT for the multi-mode (single-mode) measurements (here $\epsilon_{\rm mf} \equiv V^2/V_{\rm c,mf}^2 - 1$). The data extrapolate to a mean-field threshold at $V_{\rm c,mf}$. However, closer to the bifurcation we find $\langle\theta^2\rangle \propto \epsilon^{-\gamma}$ with $\epsilon \equiv V^2/V_c^2 - 1$, a smaller $\gamma \simeq 0.22$ ($\gamma_0 \simeq 0.5$) for the multi-mode (single-mode) measurements, and a shifted threshold $V_c > V_{\rm c,mf}$. So far as we can tell the bifurcation remains supercritical unlike the prediction for RBC.  We suggest that an explanation of this deviation from LT may be found in the nonlinear interactions of the fluctuations. Unfortunately there is as yet no detailed theory for the critical behavior of this system with which our results could be compared. 

\narrowtext

\begin{minipage}{3.12in}
\begin{table}
\caption{Parameters and results for I52 at different temperatures $\vartheta$.}
\vskip 0.1in
\begin{tabular}{cccccccc}
$\vartheta$ & $10^{9}\sigma$ & $V^2_c$ & $V^2_{\rm c,mf}$  & $\theta_0$ & $1/S_1$\\
$(^{\circ}{\rm C})$ & $(\Omega^{-1}\mbox{\rm m}^{-1})$ & $({\rm V}^2)$ & $({\rm V}^2)$  & mrad & (rad$^{-2}$)\\
\tableline
38	& 5.16	&	160.848	&	155.296	& 1.83 & 1.77\\
42	& 6.36	& 137.978	&	135.367	& 1.81 & 3.87\\
46	& 7.76	&	124.495	&	122.685	& 1.79 & 7.63\\
50	& 9.51	&	114.774	&	114.027	& 1.84 & 9.17\\
\end{tabular}
\label{tab:parameters}
\end{table}
\end{minipage}

\begin{figure}[h]
\epsfxsize=2.75in
\centerline{\epsffile{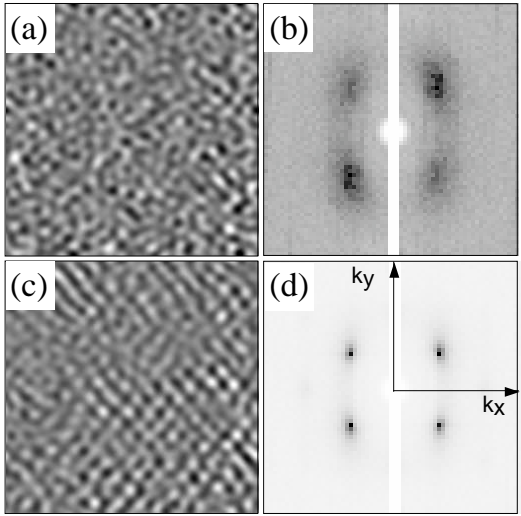}}  
\vskip 0.2in 
\caption{Background-divided and Fourier-filtered snapshots of size $502\times502~\mu$m$^2$ (a,c), and the central part $-22.4/d \leq k_x, k_y \leq 22.4/d$ of the time-averaged structure factors $S({\bf k})$ (b,d) for I52. (a,b) : $\epsilon = -0.104$. (c,d) : $\epsilon = -0.001$. The director is horizontal.}
\label{fig:Snapshots}
\end{figure}

In the experiments reported here we used samples with planar alignment. The I52 ({\it 4-ethyl-2-fluoro-4$^\prime$-[2-(trans-4-pentyl\-cyclo\-hexyl)ethyl]-bi\-phenyl}) was  doped with $4.3 \%$ by weight of molecular iodine (I$_2$) and studied in the apparatus and by the cell-assembly techniques of Ref.\ \cite{BA99}. The bifurcation to EC is supercritical over a range of the electrical conductivity $\sigma$\ \cite{DAC96a}. Above the bifurcation the pattern consists of chaotic traveling zig and zag modes.
The cell had a thickness $d = 28 \pm 1 \mu$m. We changed $\sigma$ by varying the temperature (see Tab.\ \ref{tab:parameters}) and measured $\sigma$ at a frequency of 50 Hz and $V = 2.0$ V.
For the experiment we used a frequency of 25 Hz. At each voltage we waited 110 s, and then took 128 images 10 s apart. The combination of cell thickness, I$_2$ concentration, and drive frequency assured that the ``worm" state \cite{DAC96b} did not occur for $\sigma > 4\times 10^{-9} \Omega^{-1}{\rm m}^{-1}$. For MPV we used a frequency of 30 Hz and $d = 23.1 \pm 0.2~\mu$m\ \cite{Ho96}.

Figure\ \ref{fig:Snapshots} shows two examples of single snapshots for I52 at different $\epsilon$ and for $\sigma = 7.8\times10^{-9} \Omega^{-1}$m$^{-1}$. Far below onset [Fig.\ \ref{fig:Snapshots}(a)] we see patches of patterns with a small correlation length. The patterns are very weak, although dividing by a background image, Fourier filtering,  and using the full available greyscale uncovers the expected zig and zag modes. Much closer to onset [Fig.\ \ref{fig:Snapshots}(c)] a pattern containing extended patches of zig and zag rolls is found.

\begin{figure}[h]
\epsfxsize=2.5in
\centerline{\epsffile{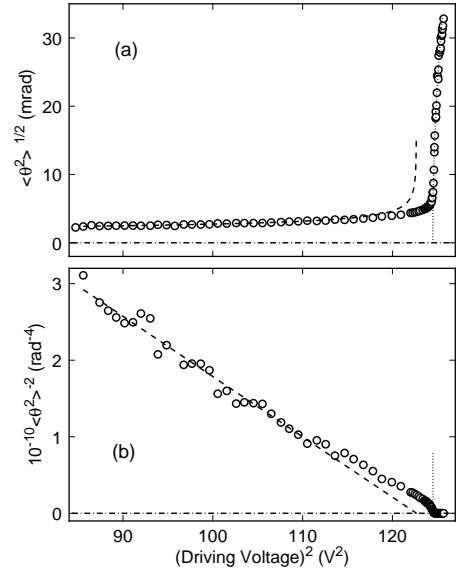}}  
\vskip 0.2in 
\caption{(a): The root-mean-square of the director-angle fluctuations $\langle\theta^2\rangle^{1/2}$ and (b): the inverse square of the mean square of the director-angle fluctuations $\langle\theta^2\rangle^{-2}$ for I52 and $\sigma = 7.8\times10^{-9} \Omega^{-1}$m$^{-1}$ as a function of the square of the driving voltage. The dotted lines are a fit of a straight line to $\langle\theta^2\rangle(V^2)$ in the range $124.53~\mbox{\rm V}^2 \leq V^2 \leq 125.15~\mbox{\rm V}^2$. They extrapolate to $V_c^2 = 124.495~\mbox{\rm V}^2$. The dashed curves are a fit of a straight line to the data in (b) in the region $85~\mbox{\rm V}^2 \leq V^2 \leq 114~\mbox{\rm V}^2$ where LT applies. In (b) the intercept of the dashed straight line  with the horizontal axis gives $V_{\rm c,mf}^2 = 122.685~\mbox{\rm V}^2$.}
\label{fig:ThetaVsVsq}
\end{figure}

For the multi-mode analysis with I52 we closely followed the method of Ref.\ \cite{BA99} to extract $\langle\theta^2\rangle$. We calculated $I_i({\bf x},\epsilon) \equiv \tilde{I}_i({\bf x},\epsilon) / \tilde{I}_0({\bf x},\epsilon) - 1$ for each image $\tilde{I}_i({\bf x},\epsilon),\ i = 1, ..., 128$. Here ${\bf x} = (x,y)$ are the coordinates in real space and $\tilde{I}_0({\bf x},\epsilon)$ is a background image obtained by averaging 128 images at the same $\epsilon$.
For each $I_i({\bf x},\epsilon)$ we derived the structure factor (the square of the modulus of the Fourier transform) $S_i({\bf k},\epsilon)$ and averaged 128 $S_i({\bf k},\epsilon)$ to get  
$S({\bf k},\epsilon)$, where ${\bf k} = (k_x,k_y)$ is the wave vector. Figures\ \ref{fig:Snapshots}(b) and (d) are examples. In agreement with Refs.\ \cite{BA98,BA99,Ho96} we see two pairs of peaks corresponding to two sets of rolls oriented obliquely to the director which get sharper and larger as we approach the onset of convection. The two modes are called zig and zag modes and correspond to those of the extended chaos above onset\ \cite{DAC96a}.
We computed the total power under the peaks of $S({\bf k})$ as described in\ \cite{BA99}, and then converted it to $\langle\theta^2\rangle$\ \cite{RHWR89} (see also Eq. (5) in\ \cite{BA99}).

Figure\ \ref{fig:ThetaVsVsq}(a) shows $\langle\theta^2\rangle^{1/2}$ as a function of $V^2$ for I52 and $\sigma = 7.8\times10^{-9} \Omega^{-1}{\rm m}^{-1}$. It is small for low voltages but then increases sharply above onset. 
As expected,\ \cite{DAC96a} the primary bifurcation is supercritical in this intermediate region of $\sigma$, and the points above onset follow 
a square-root law. The square-root fit above onset [dotted line in (a)]  extrapolates to $\langle\theta^2\rangle^{1/2} = 0$ at $V_c^2$. 

\begin{figure}[h]
\epsfxsize=2.5in
\centerline{\epsffile{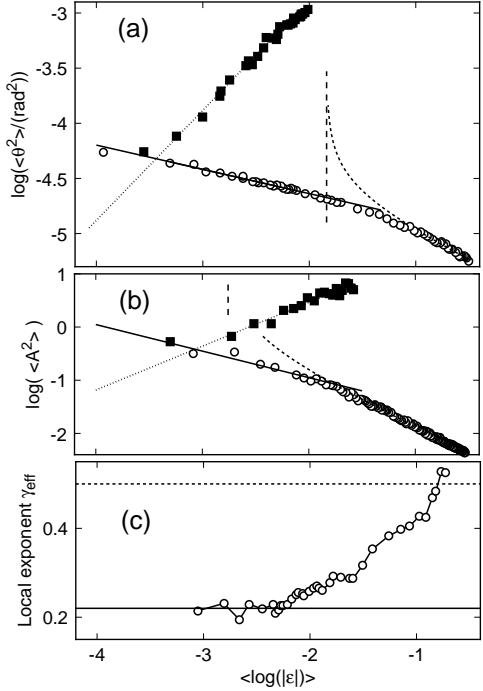}}  
\vskip 0.2in 
\caption{(a): Mean square $\langle\theta^2\rangle$ for I52, averaged over all modes and over 128 images, of the director angle as a function of $\log(|\epsilon|)$ (open circles: $\epsilon < 0$, solid squares: $\epsilon > 0$). Short dashed curve: fit to data in the linear region from Fig.~\protect{\ref{fig:ThetaVsVsq}(b)}. Straight solid line: fit for $\epsilon < 0$ in the range $-3.31 \leq \log(|\epsilon|) \leq -2.13$. It yields $\gamma = 0.22$. Dotted line: fit to data above onset from Fig.~\protect{\ref{fig:ThetaVsVsq}(a)}. Vertical dashed line: location of $V_{\rm c,mf}$ along the $\epsilon$-axis. (b): Mean square $\langle A^2 \rangle$ of single-mode fluctuation amplitudes (arbitrary scale) for MPV as a function of $\log(|\epsilon|)$. The lines and symbols have meanings equivalent to those in (a). (c): Effective local exponent $\gamma_{\rm eff}$ of $\langle\theta^2\rangle$ in (a) for $\epsilon < 0$. It was obtained from fits within a sliding window covering half a decade of $\epsilon$. Dashed horizontal line: $\gamma = 0.5$. Solid horizontal line:  $\gamma =  0.22$ as suggested by our data at small $|\epsilon|$.}
\label{fig:ThetaVsEpsilon}
\end{figure}

In Fig.\ \ref{fig:ThetaVsVsq}(b) we plotted $\langle\theta^2\rangle^{-2}$ versus $V^2$. The data are the same as those in Fig.\ \ref{fig:ThetaVsVsq}(a). For $V^2 \leq 113.6~\mbox{\rm V}^2$ they can be fitted by a straight line. This is in agreement with LT, which gives  $\langle \theta^2 \rangle  = \theta_0^2\epsilon_{\rm mf}^{-1/2}$. The straight-line fit shown in the linear  region of Fig.\ \ref{fig:ThetaVsVsq}(b) yielded $V_{\rm c,mf}^2 = 122.685~\mbox{\rm V}^2$ and $\theta_0 = 1.8$ mrad. The value of $\theta_0$ is in fairly good agreement with the approximate theoretical estimate $\theta_0 = 2.8$ mrad (see Ref.\ \cite{RRDSHAB91} and Eq. (6) in\ \cite{BA99}). Hence, the measured mean-square director-angle fluctuations are consistent with thermal noise. The mean-field threshold $V_{\rm c,mf}^2$ is well below the actual bifurcation point at $V_{\rm c}^2$. A similar, albeit somewhat smaller, shift was found for MPV (see Fig.\ \ref{fig:ThetaVsEpsilon}(b) below).

Figure\ \ref{fig:ThetaVsVsq}(b) reveals that the points close to but below onset deviate from the prediction of LT. Similar deviations were  observed over a range of $\sigma$ (see Tab.~\ref{tab:parameters}), as well as for MPV. To illustrate the behavior close to the bifurcation, we plotted $\log(\langle \theta^2 \rangle)$ versus $\log(|\epsilon|)$ in Fig.\ \ref{fig:ThetaVsEpsilon}(a) for I52 and $\sigma = 7.8\times10^{-9} \Omega^{-1}$m$^{-1}$. For $\epsilon < 0$ (open circles) the data points far away from $V_c$ (far right of the figure) follow the behavior predicted by LT (short-dashed curve). Note that in a plot where the horizontal axis is  $\log(|\epsilon_{\rm mf}|)$ this should give a straight line with a slope of -1/2 for the multi-mode analysis used here. The line is curved and diverges at the $\epsilon$-value corresponding to $V^2_{\rm c,mf}$. Closer to onset the data points deviate from the linear prediction. We find a crossover to a regime which can be described by a new powerlaw: $\langle \theta^2 \rangle \propto \epsilon^{-\gamma}$ with $\gamma \simeq 0.22$. In order to characterize this crossover,  we calculated the local exponent $\gamma_{\rm eff}(\epsilon)$ from fits to data spanning half a decade of $\epsilon$. Figure\ \ref{fig:ThetaVsEpsilon}(c) shows that $\gamma_{\rm eff}$ matches the expected $\gamma = 0.5$ in the linear region at large $|\epsilon|$. As we approach the onset of convection, the exponent approaches a plateau of 0.22. Thus we plotted a line in Fig.\ \ref{fig:ThetaVsEpsilon}(a) with the slope of  -0.22 (straight solid line) and found good agreement with the data over a wide range.

\begin{figure}[h]
\epsfxsize=2.5in
\centerline{\epsffile{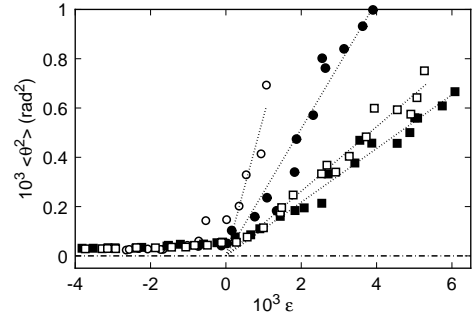}}  
\vskip 0.2in 
\caption{Mean square $\langle\theta^2\rangle$ of the director angle $\theta$  close to onset as a function of $\epsilon$ for different $\sigma$. Straight dotted lines are fits above onset. Open circles, solid circles, open squares, and solid squares correspond to $\sigma = 5.2, 6.4, 7.8, $ and $9.5 \times 10^{-9} \Omega^{-1}{\rm m}^{-1}$, respectively.}
\label{fig:OnsetDifferentSigma}
\end{figure}

Data for MPV are shown for comparison in Fig.\ \ref{fig:ThetaVsEpsilon}(b). As mentioned above, they are based on a single-mode analysis. Far below the bifurcation they yield the expected LT exponent $\gamma_0 = 1$ (note that the resolutions of the vertical scales of Figs.\ \ref{fig:ThetaVsEpsilon}(a) and (b) differ by a factor of two so that the corresponding slopes look equal). With decreasing $|\epsilon|$ there is a crossover to a smaller exponent close to $\gamma_0 = 0.5$, which is  consistent with the crossover from $\gamma = 0.5$ to $\gamma = 0.22$ found from the multimode-analysis for I52. 

\begin{figure}[h]
\epsfxsize=2.5in
\centerline{\epsffile{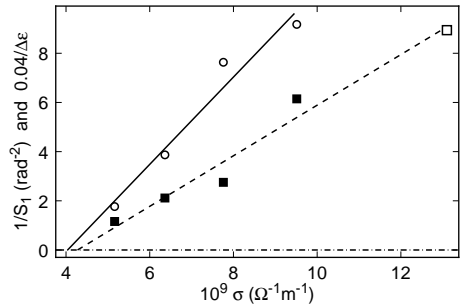}}  
\vskip 0.2in 
\caption{The inverses of the initial slope $S_1$ (open circles) of $\langle\theta^2\rangle$ above onset, and of the onset shift $\Delta \epsilon$ (solid squares: this work, open square: from Ref.~\protect{\cite{BA99}}) as a function of $\sigma$. The solid line is a fit to $1/S_1$. Its intercept with $1/S_1 = 0$  suggests a tricritical point at $\sigma_t = 4.0\times10^{-9} \Omega^{-1}{\rm m}^{-1}$. A straight-line fit to $1/\Delta \epsilon$ also passes through  zero near  $\sigma_t$.}
\label{fig:GAndDeltaEpsilon}
\end{figure}

For I52, measurements similar to those discussed above were made at four different conductivities (see Tab.\ \ref{tab:parameters}). Figure\ \ref{fig:OnsetDifferentSigma} shows the results in the vicinity of onset. The data for $\epsilon > 0$ can be fit by $\langle \theta^2 \rangle = S_1(\sigma) \epsilon$ as expected for a supercritical bifurcation. The slope $S_1$ increases as $\sigma$ decreases. 
In Fig.\ \ref{fig:GAndDeltaEpsilon} we show $1/S_1$ (open circles) as a function of $\sigma$. The data suggest that
 $1/S_1$ vanishes at $\sigma_t \simeq 4.0\times10^{-9} \Omega^{-1}$m$^{-1}$.  This is consistent with a tricritical bifurcation at $\sigma_t$, with a transition from a supercritical to a subcritical bifurcation as $\sigma$ decreases below $\sigma_t$.
The results for $1/S_1$ are consistent with recent weakly-nonlinear calculations\ \cite{TK98}.

In Fig.~\ref{fig:GAndDeltaEpsilon} we also show $1/\Delta\epsilon$ where $\Delta\epsilon \equiv (V_c - V_{\rm c,mf})/V_c$. Here the solid squares are from the present work, and the open square is from\ \cite{BA99}\ \cite{FN1}. The dashed straight line is a least-squares fit to the data. Within our resolution it passes through zero at $\sigma_t$, suggesting that the critical region broadens strongly near $\sigma_t$\ \cite{FN2}.

In this paper we showed that the fluctuations below onset of electroconvection reveal a crossover as the bifurcation is approached from the behavior predicted by linear theory to a different region characterized by a smaller exponent and a shifted threshold. In analogy to equilibrium phase transitions, we suggest that this phenomenon may be attributable to nonlinear interactions between the fluctuations. Equivalent behavior was found in the two NLCs I52 and MPV, even though we used very different methods of analysis for the two cases.   

The work on I52 was done at Santa Barbara and supported by National Science Foundation Grant DMR00-71328. The work on MPV was supported by Deutsche Forschungsgemeinschaft Grant Re-588/6. One of us (MAS) greatfully acknowledges support through a Feodor Lynen Fellowship from the Alexander von Humboldt-Foundation.

\end{multicols}


\begin{references}
\bibitem{FN0} The large susceptibility to external noise of NLCs was recognized  by R. Graham (see Sect. 4.2 of Ref. \cite{Gr75}).
\bibitem{FN1} We re-analyzed the data in Fig. 4 of Ref.~\cite{BA99} by fitting those for $V^2 < 0.9 V_c^2$ to a straight line. This gave $V_{\rm c,mf} = 103.366~\mbox{\rm V}^2$ and $\epsilon_{\rm mf} = 0.0045 $.
\bibitem{FN2}We note that such a broadening could possibly be caused by a strong increase of $V_c$ with decreasing $\sigma$, i.e., with a strongly decreasing angle between the experimental path of constant $\sigma$ and the bifurcation line. However, the data for $V_c$ given in Tab.~\ref{tab:parameters}  show that $V_c$ remains finite near $\sigma_t$, indicating that path renormalization is unlikely to be responsible for the strong increase of $S_1$ and $\Delta \epsilon$ at $\sigma_t$.
\end{references}
\end{document}